\documentclass[twocolumn,prb,aps,longbibliography,
superscriptaddress,10pt]{revtex4-1}

\usepackage{amsfonts,amssymb}
\usepackage[sumlimits,intlimits]{amsmath}
\usepackage{graphics}
\usepackage{graphicx}
\usepackage{mathrsfs}
\usepackage{textcomp}
\usepackage{verbatim}
\usepackage{hyperref}    % links
\usepackage{bm}          % bold math letters

\newcommand{\be}{\begin{equation}}
\newcommand{\ee}{\end{equation}}
\newcommand{\e}{\varepsilon}

\newcommand{\TES}{T_\text{ES}}

\begin{document}

%\title{Fractal structure of hopping current in completely compensated semiconductors \\
%OR \\
\title{Enhancement of hopping conductivity \\ by spontaneous fractal ordering of low-energy sites}
%\title{Is there Efros-Shklovskii conductivity outside the Coulomb gap?}

\author{Tianran Chen}
\affiliation{West Chester University, West Chester, PA 19383  USA}

\author{Brian Skinner}
\affiliation{Massachusetts Institute of Technology, Cambridge, MA 02139  USA}

\date{\today}

\begin{abstract}

Variable-range hopping conductivity has long been understood in terms of a canonical prescription for relating the single-particle density of states to the temperature-dependent conductivity.  Here we demonstrate that this prescription breaks down in situations where a large and long-ranged random potential develops.  In particular, we examine a canonical model of a completely compensated semiconductor, and we show that at low temperatures hopping proceeds along self-organized, low-dimensional subspaces having fractal dimension $d = 2$.  We derive and study numerically the spatial structure of these subspaces, as well as the conductivity and density of states that result from them.  One of our prominent findings is that fractal ordering of low energy sites greatly enhances the hopping conductivity, and allows Efros-Shklovskii type conductivity to persist up to unexpectedly high temperatures.

\end{abstract}

\maketitle

\section{Introduction}
\label{sec:intro}

In disordered systems with localized electrons, there is no electronic conductivity at zero temperature.  At nonzero temperature, however, a finite conductivity is enabled by the process of phonon-assisted quantum tunneling, or ``hopping", of electrons between localized states.  When the temperature $T$ is relatively high, phonons are abundant and electrons hop primarily between nearest-neighboring sites, with an activation energy that is given by the average difference in energy between neighboring sites.  On the other hand, when the temperature is low, nearest-neighbor hopping events become extremely rare and the conductivity is dominated by long-distance hops, in which an electron tunnels directly to a distant target site with energy closely-matched to that of its initial site.  This process is known as variable-range hopping (VRH), and was first described by Mott nearly 50 years ago \cite{mott_conduction_1968}.  

Mott's initial theory was eventually generalized into a prescription for determining the temperature-dependent conductivity, $\sigma(T)$, from the energy-dependent single-particle density of states (DOS), $g(\e)$.  (Here and below, $\e$ is defined relative to the Fermi level.)  In this prescription, one first considers that at a fixed temperature $T$ electrons hop within some band of energies $\e \in (-\e_0, \e_0)$ around the Fermi level.\cite{shklovskii_electronic_1984}  The typical spatial concentration of sites within this band is $N(\e_0) = \int_{-\e_0}^{\e_0} g(\e) d\e$ and the typical hopping distance can therefore be estimated as $r(\e_0) = [N(\e_0)]^{-1/d}$, where $d$ is the system dimensionality.  The temperature-dependent conductivity is then calculated by maximizing the typical electron tunneling rate
\be 
\Gamma \propto \exp\left[- \frac{2 r(\e_0)}{\xi} - \frac{\e_0}{T} \right]
\label{eq:gamma}
\ee 
with respect to $\e_0$, where $\xi$ is the localization length and $T$ is taken to have units of energy.  In Eq.\ (\ref{eq:gamma}), the first term in the exponential indicates suppression of tunneling with distance due to exponentially small overlap between electron wavefunctions, and the second term indicates suppression with energy due to vanishing phonon abundance.  The value of $\Gamma$ is proportional to $\sigma$ to within algebraic prefactors.  This prescription for determining $\sigma(T)$ from $g(\e)$ can be called the ``Mott doctrine" for understanding VRH, and it has been at the heart of our understanding of hopping conduction for the last half-century.  Most generally, if the single-particle DOS varies with energy according to $g(\e) \propto |\e|^{\alpha}$, then the Mott doctrine gives
\be 
\ln \sigma \propto T^{- (\alpha + 1)/(d + \alpha + 1)}.
\label{eq:Mottdoctrine}
\ee 

\begin{figure}[htb]
\centering
\includegraphics[width=0.5 \textwidth]{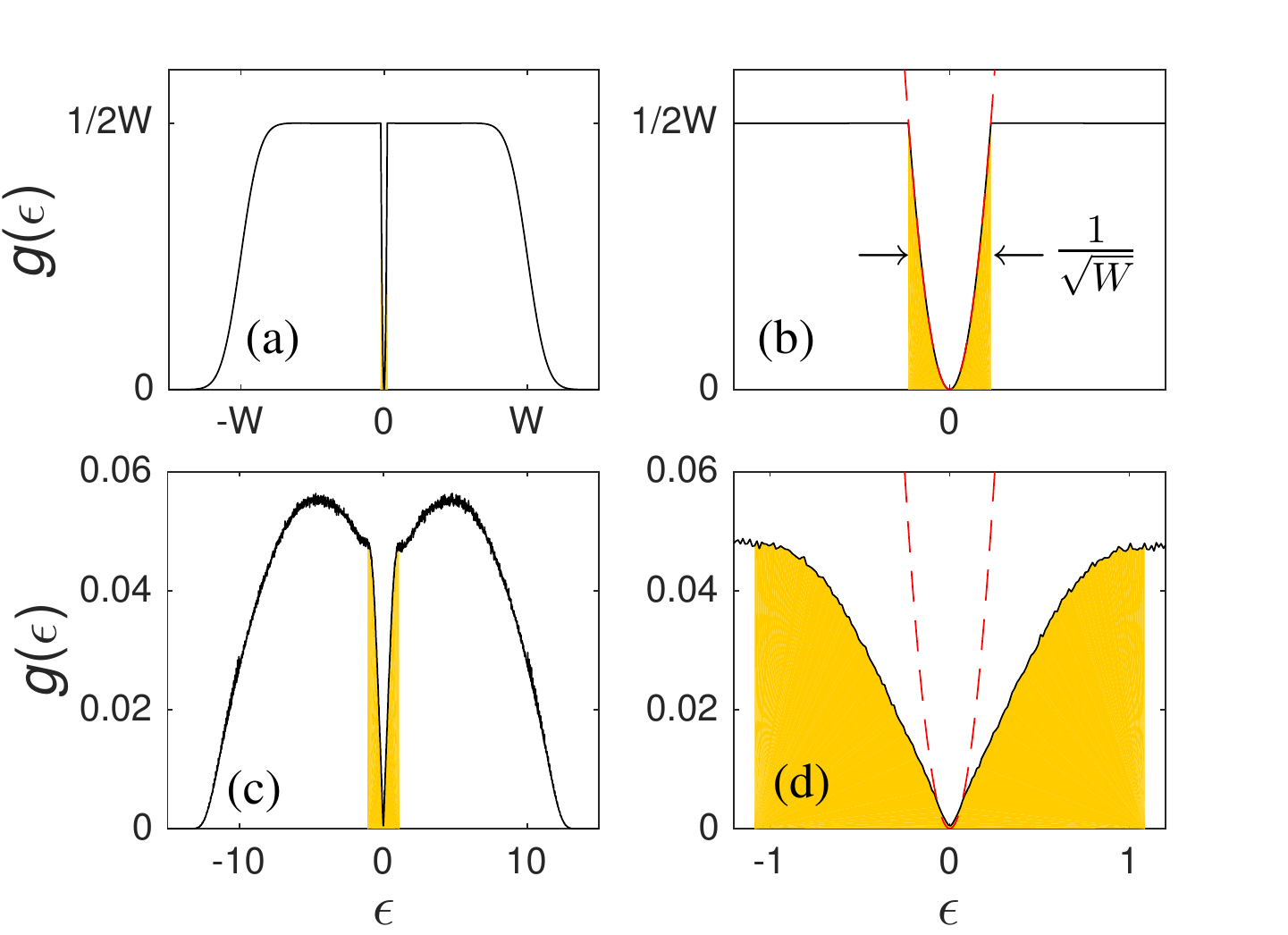}
\caption{Single-particle DOS as a function of energy.   (a) and (b) show the typical situation, where electron-hole correlations arising from the ES stability criterion dig a narrow Coulomb gap out of a roughly constant DOS. If the total width of the DOS is $\sim 2W$, then the Coulomb gap width is $\sim 1/\sqrt{W}$. (c) and (d) show the results from our numeric simulation for $\Delta = 10$.  The red dashed line shows the 3D Coulomb gap, Eq.\ (\ref{eq:CG}).  Shaded regions in these plots show the band of energies, as measured by our numeric simulations, that are used at the highest temperature at which we observe ES VRH [Eq.\ (\ref{eq:ES})].  Units of energy in this plot are $e^2/a_0$ and $g(\e)$ is plotted in units of $(e^2 a_0^2)^{-1}$.}
\label{fig:DOScompare}
\end{figure}

While Mott initially assumed a constant DOS around the Fermi level ($\alpha = 0$), it was later shown by Efros and Shklovskii \cite{efros_coulomb_1975} (ES) (after earlier suggestions by Pollak,\cite{pollak_effect_1970} Knotek,\cite{knotek_correlation_1974} and Srinivasan\cite{srinivasan_statistical_1971}) that interaction-induced correlations between electrons generically guarantee a vanishing DOS at $\e \rightarrow 0$.  One can understand this vanishing DOS by noticing that, in the ground state arrangement, a filled site and an empty site cannot simultaneously have an energy difference $\e$ and a spatial separation $r < e^2/\e$.  If they did, then the system would be unstable with respect to the process of removing the electron from the filled site (upon which the energy of the empty site would drop, since the contribution to the Coulomb potential by the electron at the formerly-filled site would vanish) and adding it to the empty site.  In other words, a filled site having energy $\e_\text{f}$ and an empty site having energy $\e_\text{e}$ must satisfy
\be 
\e_\text{e} - \e_\text{f} > e^2/r_\text{ef},
\label{eq:EScrit}
\ee 
where $r_\text{ef}$ is the distance between the two sites.  (Here, $e^2$ denotes the squared electron charge divided by the dielectric constant, in cgs units.)  Equation (\ref{eq:EScrit}) is usually referred to as the ``ES stability criterion".  It guarantees that sites which are close together in energy cannot be close together in space, and therefore that the total concentration of sites with energy smaller than $\e$ is limited to $N \lesssim (e^2/\e)^{-d}$.  Defining the DOS $g = dN/d\e$ gives \cite{efros_coulomb_1976}
\be 
g(\e) \lesssim \frac{|\e|^{d-1}}{e^{2d}}.
\label{eq:CG}
\ee 
Equation (\ref{eq:CG}) is called the ``Coulomb gap"; its typical effect on the DOS in a three-dimensional (3D) system is illustrated in Fig.\ \ref{fig:DOScompare}(a)-(b).
Using the ideas of the Mott doctrine outlined above [Eq.\ (\ref{eq:Mottdoctrine})], the conductivity resulting from the Coulomb gap is
\be 
\sigma \propto \exp\left[ - (\TES/T)^{1/2} \right],
\label{eq:ES}
\ee 
regardless of the system dimensionality, where $\TES \sim e^2/\xi$. Equation (\ref{eq:ES}) is called the ES law of conductivity, and it arises generically in disordered, localized systems at sufficiently low temperature.  The universality of Eqs.\ (\ref{eq:CG}) and (\ref{eq:ES}) have inspired a large number of analytical, numerical, and experimental studies during the past four decades. (For a limited selection of these, see, for example, Refs.\ \onlinecite{baranovskii_coulomb_1979, larkin_activation_1982, davies_properties_1984, mobius_coulomb_1992, pikus_critical_1994, li_unexpected_1994, yu_time-dependent_1999, lee_coulomb_1999, butko_coulomb_2000, muller_glass_2004, surer_density_2009, goethe_phase_2009, amir_variable_2009,  bardalen_coulomb_2012}).

Importantly, the ES law is expected to apply only at sufficiently low temperatures that the band of energies used for hopping conductivity is within the range where Eq.\ (\ref{eq:CG}) applies.  In other words, ES conductivity is supposed to occur only when the states used for conduction are inside the Coulomb gap.  Consider, for example, the most common situation, where electrons hop in a three-dimensional (3D) system for which the single-particle energies fall within some wide range $\e \in (-W, W)$.  In this situation the ``bare DOS" is determined by the total width of the energy band, $g_0 \sim 1/(a_0^3 W)$, where $a_0$ is the typical distance between nearest-neighboring sites.  The DOS is expected to follow the Coulomb gap expression, $g_\text{CG}(\e) \sim \e^2/e^6$, only at such small energies that $g_\text{CG}(\e) < g_0$.  This condition implies that the Coulomb gap has a width $\e_\text{CG} \sim \sqrt{e^6/a_0^3 W}$ that becomes increasingly narrow as the range $W$ of site energies is increased.  Consequently, the ES law, Eq.\ (\ref{eq:ES}), describes the conductivity only at low temperatures $T \lesssim e^4 \xi/(a_0^3 W)$.  This situation is illustrated schematically in Fig.\ \ref{fig:DOScompare}(a)-(b).

In this paper we demonstrate that this canonical expectation of ES conductivity inside a parabolic Coulomb gap can in fact be strongly violated.  We examine a simple model that describes hopping transport in a completely compensated semiconductor, and we demonstrate that for this system the Mott doctrine itself is apparently violated, so that the usual relation between $g(\e)$ and $\sigma(T)$ is broken.  The result of this breakdown is that the regime of validity of Eq.\ (\ref{eq:ES}) is enormously extended in temperature, and ES conductivity persists far outside the regime of a quadratic Coulomb gap [as illustrated in Fig.\ \ref{fig:DOScompare}(c)-(d)].  We explain this surprising result in terms of a spontaneous fractal ordering of low-energy sites, such that for hopping at low temperature the dimensionality $d$ of the system is effectively reduced.  Our arguments constitute an update to the theory of transport in completely compensated semiconductors, and can be used to resolve prominent puzzles about the hopping transport in such diverse systems as nanocrystal arrays  \cite{liu_mott_2010, skinner_theory_2012} and the bulk transport of doped, compensated topological insulators \cite{ren_optimizing_2011, skinner_why_2012}, where an anomalous prolongation of ES conductivity to high temperatures was observed but never explained.

The remainder of this paper is organized as follows.  In Sec.\ \ref{sec:model} we define the model to be studied and discuss our numerical simulations.  In Sec.\ \ref{sec:DOS} we present scaling arguments and numerical results to identify the fractal spatial structure of low-energy sites, and we give results for the DOS.  Section \ref{sec:conductivity} uses these results to derive the temperature-dependent hopping conductivity, which we verify with simulation results.  We conclude in Sec.\ \ref{sec:conclusion} with a summary and discussions about relevant recent experiments and the generality of our results.

\section{Model and numerical simulations}
\label{sec:model}

\subsection{Model}

The model we consider can be viewed as a lattice-discretized version of a canonical model of completely compensated semiconductors \cite{shklovskii_completely_1972, shklovskii_electronic_1984}, in which electrons can reside on ``donor'' or ``acceptor'' sites that occur in equal number.  We arrange these sites into a cubic lattice, with lattice constant $a_0$ and size $L$ in each orthogonal direction.  Each site is randomly assigned to be either a donor or acceptor type and can be occupied by at most one electron.  Donor and acceptor sites are associated with on-site energies $\pm \Delta/2$, respectively, and electrons interact via an unscreened Coulomb interaction.  The corresponding (classical) Hamiltonian is
\be 
H = \frac{\Delta}{2} \sum_{i} f_i n_i + \sum_{ij} \frac{e^2 }{r_{ij}}q_i q_j,
\ee
where $f_i$ is a binary variable that discriminates between donors ($f_i = 1$) and acceptors ($f_i = -1$), and $n_i = 0, 1$ is the electron occupation of the $i$th lattice site.  Donor sites are neutral when occupied and acceptors are neutral when empty, so that $q_i = (f_i + 1)/2 - n_i$ is the charge of site $i$.  

This model differs notably from the most well-studied model for hopping transport, in which the term $(\Delta/2)f_i$ is replaced by a random on-site energy that is drawn from a wide, continuous distribution (as studied, for example, in Refs.\ \onlinecite{efros_coulomb_1976, mobius_coulomb_1992, goethe_phase_2009}).  As we discuss below, in our model the binary nature of the on-site energy implies a prominent role for the long-ranged Coulomb potential created by electrons and donor charges.  

For the remainder of this paper we take units where $e^2 = a_0 = 1$, so that all lengths are written in units of $a_0$ and all energies are in units of $e^2/a_0$.  Thus, the only physical parameter of the model is the band gap energy $\Delta$.  Our focus in this paper is on the limit $\Delta \gg 1$.

\subsection{Pseudo-ground states}

We study this system numerically using a computer simulation, in which we search for the ground state electron occupation $n_i$ of each lattice site for a given random assignment of each $f_i$ and a given choice of $\Delta$ (details of the search algorithm are given, \textit{e.g.}, in Ref.\ \onlinecite{shklovskii_electronic_1984}).  We operate our simulation in the grand-canonical ensemble with fixed chemical potential $\e = 0$, so that in the ground state all sites with $\e < 0$ are filled and all sites with $\e > 0$ are empty.  

Of course, finding the exact ground state for a system with reasonable size is a very difficult computational problem, and in general our algorithm does not produce the exact ground state of the system, but only a ``pseudo-ground state" that is stable with respect to the simultaneous change of any two electron occupation numbers [\textit{i.e.,} Eq.\ (\ref{eq:EScrit}) is satisfied for all possible choices of filled and empty sites].  Higher-order stability criteria have been studied elsewhere (for example, in Refs.\ \onlinecite{goethe_phase_2009, efros_coulomb_2011}), and their effects are generally very small in the range of energies that we consider in this paper.  

Once the set of pseudo-ground state occupation numbers, $\{n_i\}$, are known, we calculate the single-electron energy $\e_i$ at each lattice site:
\be 
\e_i = \frac{\Delta}{2}f_i - \sum_{j \neq i} \frac{q_j}{r_{ij}}.
\ee 
The DOS is calculated by making a histogram of the values of $\e_i$.  Results presented below for the DOS correspond to simulated systems of size $L = 80$.

\subsection{VRH conductivity}

Once the pseudo-ground state energy at each site is known, we calculate the conductivity of the system using the approach of the Miller-Abrahams resistor network \cite{miller_impurity_1960}.  This process is described in detail elsewhere (for example, in Refs.\ \onlinecite{shklovskii_electronic_1984, efros_coulomb_1985}); briefly, it involves modeling the tunneling rate between a given pair of (not necessarily neighboring) sites using a resistance whose value increases exponentially with the sites' spatial separation and with the corresponding activation energy [as suggested by Eq.\ (\ref{eq:gamma})].  The conductivity of the network is approximated by identifying the minimum value of the resistance such that, when all higher-resistance links are abandoned, current can still percolate through the network.

Since the magnitude of the conductivity at a given temperature depends on the value of the localization length $\xi$, we find it expedient to define dimensionless units that incorporate $\xi$ in such a way that it is eliminated as a free variable.  In particular, notice that at a given temperature the conductivity can be written as
\be 
\sigma = \sigma_0 \exp\left[ -\frac{2 r}{\xi} - \frac{\e_a}{T} \right],
\label{eq:sigmarT}
\ee 
where $\e_a$ is the activation energy, $r$ is the typical hopping distance, and $\sigma_0$ is a prefactor the depends only algebraically on temperature.  We therefore introduce the dimensionless temperature (restoring explicitly the units $e^2$ and $a_0$)
\be 
T^* = \frac{2 a_0^2 T}{e^2 \xi}
\label{eq:Tstar}
\ee
and the dimensionless logarithm of the conductivity:
\be 
(\ln \sigma)^* = \frac{\xi}{2 a_0} \ln(\sigma/\sigma_0).
\ee 
In these units Eq.\ (\ref{eq:sigmarT}) can be written simply as
\be 
(\ln \sigma)^* = -r - \e_a/T^*
\label{eq:sigmarTdimensionless}
\ee 
(reverting to units where $e^2 = a_0 = 1$), and $\xi$ is eliminated as a separate variable for describing the conductivity.

\subsection{Nearest-neighbor hopping conductivity}
\label{sec:NNH}

The resistor network method for calculating conductivity is expensive computationally, since it involves calculating the effective resistance between all pairs of sites in the system for each value of the temperature.  We are therefore limited to studying VRH numerically at only modestly large system sizes $L = 30$.  For reasons discussed in Sec.\ \ref{sec:DOS}, this finite size limitation also prevents us from applying the resistor network method to values of $\Delta$ larger than $10$.  Nonetheless, the high temperature conductivity (which we discuss in Sec.\ \ref{sec:conductivity}) can be calculated using a simpler method.  At high temperatures, VRH is abandoned in favor of nearest-neighbor hopping with a finite, temperature-independent activation energy $\e_a$.  The value of $\e_a$ can be found by isolating the set of sites whose energies fall within some range $\e \in (-\e_0, \e_0)$ and checking whether those sites produce nearest-neighbor percolation across the system.  The smallest value of $\e_0$ for which such percolation is possible approximates the activation energy $\e_a$.  Using this method we can estimate the high-temperature activation energy for systems as large as $L = 80$ and $\Delta$ as large as $20$.

\subsection{Fractal dimension}
\label{sec:fractalnumerics}

Crucial to our explanation of the nature of hopping conductivity in this system is the existence of a nontrivial fractal dimension (or ``correlation dimension") for different sets of low energy sites.  In principle, this fractal dimension can be measured directly from our numeric simulations by the method of calculating a correlation integral for a given set of sites.  This technique was introduced in Ref.\ \onlinecite{grassberger_measuring_1983}, and it can be described heuristically as follows.  First, one identifies the set of sites whose energy lies within a given energy window.  One can then choose an arbitrary site within the set, and define a sphere of radius $s$ centered at the chosen site.  The number of other sites in the set whose spatial location is also inside the sphere is $M(s)$.  The proportionality $M \propto s^d$ defines the fractal dimension $d$.  For the results presented below, we average the value of $M(s)$ over all choices of the chosen center site and over many random realizations of the donor and acceptor assignments.

\section{DOS and Structure of the random potential}
\label{sec:DOS}

\subsection{Long-ranged potential}

\begin{figure}[htb]
\centering
\includegraphics[width=0.48 \textwidth]{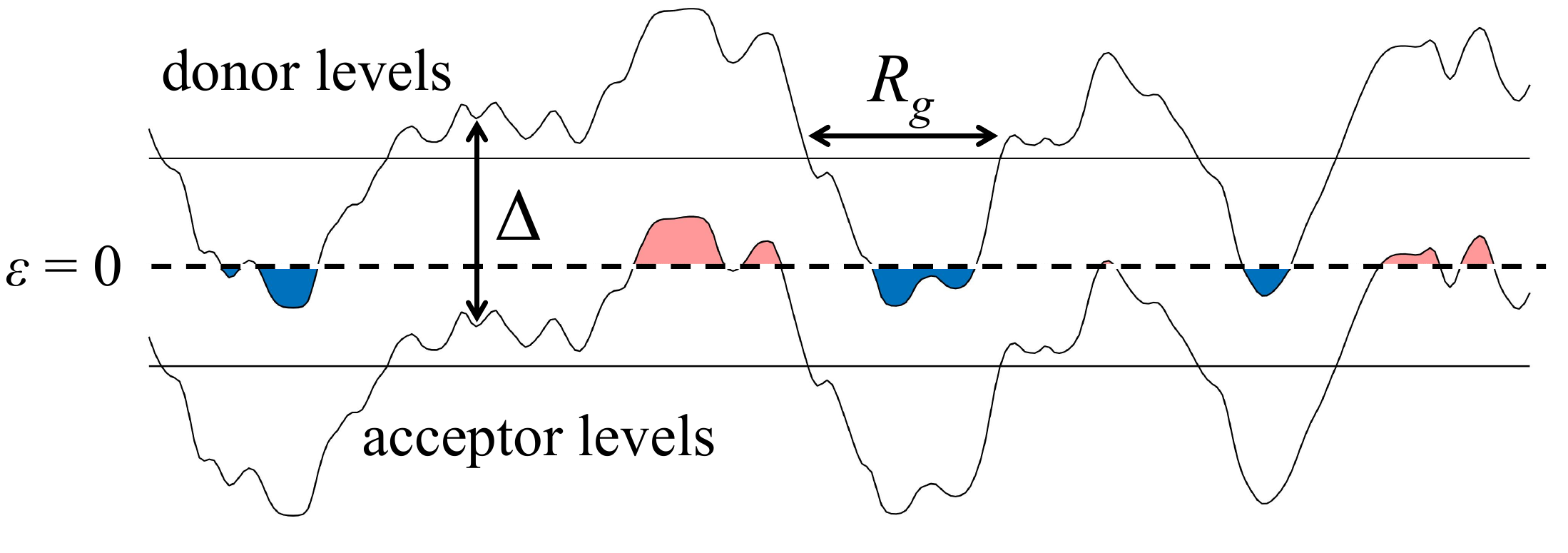}
\caption{Schematic illustration of the random potential.  The upper (lower) meandering thin lines show the energy of donor (acceptor) sites as a function of spatial coordinate. Screening occurs when these levels cross the Fermi level ($\e = 0$), at which point donor or acceptor sites become neutralized (indicated by red and blue shaded areas, respectively) and further growth of the potential is truncated. $R_g$ denotes the correlation length of the potential, and horizontal thin lines show $\e = \pm \Delta/2$.}
\label{fig:bandbending}
\end{figure}

In order to understand the transport in this system, and the apparent failure of the Mott doctrine, one must first consider the structure of the random Coulomb potential.  As is generically the case in heavily compensated semiconductors, in our model a large value of $\Delta \gg 1$ implies that a large, random Coulomb potential develops throughout the system \cite{shklovskii_completely_1972, shklovskii_electronic_1984, skinner_effects_2013}.  One can see the origin of this potential by first imagining that all donors (with nominal energy $+\Delta/2$) are empty, while all acceptors (with nominal energy $-\Delta/2$) are filled.  In this situation all sites become charged either positively (for donors) or negatively (for acceptors), and consequently a random Coulomb potential develops.  While the system is neutral on average, statistical fluctuations in the local concentration of donors or acceptors lead to large fluctuations in the Coulomb potential over long length scales, which can be understood using the following scaling argument\cite{shklovskii_completely_1972}.   Consider a Gaussian sphere of size $R \gg 1$.  Such a sphere encloses $\sim R^3$ charges, which sum to zero on average but whose sum has a typical magnitude $Q \sim R^{3/2}$.  The typical potential at the surface of the sphere, relative to the value at its center, is $\phi(R) \sim Q/R \sim \sqrt{R}$.  One can thus think that the potential experiences increasingly large fluctuations over increasingly long length scales, as in a random walk (illustrated in Fig.\ \ref{fig:bandbending}).

These fluctuations grow with length scale until some typical length $R_g$, at which the potential becomes large enough to bring donor or acceptor levels to the Fermi level. Further growth of the potential is truncated, since donors or acceptors that cross the Fermi level change their charge state (from $q = \pm 1$ to $q = 0$) and thereby screen the potential. The resulting range of energies for electron sites is therefore as wide as $\sim (-\Delta, \Delta)$.  The length $R_g$ can be estimated by equating $\phi(R_g) \sim \Delta$, which gives (with proper numerical coefficients \cite{shklovskii_completely_1972, shklovskii_electronic_1984})
\be 
R_g = \frac{\Delta^2}{8 \pi}.
\label{eq:Rg}
\ee
$R_g$ can be thought of as the correlation length of the random potential.

The wide range of single-particle energies is shown in Fig.\ \ref{fig:DOScompare}(c), where we plot the DOS for $\Delta = 10$ as measured by our computer simulation.  
One can notice from Fig.\ \ref{fig:DOScompare}(d) that the DOS follows the 3D Coulomb gap expression, Eq.\ (\ref{eq:CG}), only over a very narrow range of energies: $|\e| \lesssim 0.06$. At larger energies $0.06 \lesssim |\e| \lesssim 1$ the DOS is better described by $g(\e) \propto |\e|$.  This unusual linear dependence of $g(\e)$ is one of the central puzzles to be explained.

\subsection{Fractal dimension}

Given this picture of a random potential with long-ranged spatial correlations, as illustrated in Fig.\ \ref{fig:bandbending}, one can understand intuitively why the Mott doctrine should fail.  Implicit in the Mott doctrine is the assumption that energy levels are drawn randomly from some distribution with no spatial structure.  Thus, in the Mott doctrine one estimates the typical distance between sites within some low-energy band in a mean-field way: by calculating the volume-averaged concentration of such sites and raising this concentration to the power $-1/d$.  However, in the present problem the gradual development of the random potential implies that sites which are close together in space are more likely than random to be close together in energy.  In other words, if two (say) donor sites have a spatial separation $r \ll R_g$, then it is unlikely that they can have a difference in energy that is of order $\Delta$.  One can thus expect that sites with similar energy are spatially clustered in such a way that their effective dimension $d$ is smaller than $3$.  

The value of $d$ can be derived theoretically using the following scaling arguments.  In these arguments, our primary goal is to derive the fractal dimension of sites with energy close to the Fermi level.  First, however, we find it expedient to derive the fractal dimension of sites with energy \emph{far} from the Fermi level.  This derivation provides a useful intermediate result which we then use to derive the fractal dimension of low-energy sites.

\subsubsection{Fractal dimension of sites with energy far from zero}

In order to understand the clustering of sites into fractal subspaces, let us first consider the set of sites having energy far from the Fermi level.  For example, take the set of sites with energy in the infinitesimal window $(\Delta/2, \Delta/2 + d\e)$; such sites are donors experiencing an electric potential close to zero.  (One could likewise consider the equivalent set of acceptor sites with energy close to $-\Delta/2$.)  Imagine now the process of drawing a box of size $s$ around a given site within this set, and let $M(s)$ denote the expected number of additional sites within the set that will also be inside the box.  In a system with no spatial correlations, one would estimate the number of sites within the box to be given by the volume-averaged concentration of sites within the set multiplied by the box volume: $M(s) \sim g(\Delta/2) d\e s^3$.  This result is indeed correct at distances $s \gg R_g$, where the correlated structure of the potential is lost due to screening effects.  Substituting $g(\Delta/2) \sim 1/\Delta$ therefore gives $M(s) \sim s^3/\Delta$ at $s \gg R_g$.

On the other hand, at distances $s \ll R_g$, there is no screening, and the structure of the random potential produces a nontrivial fractal dimension $d$ of sites, $M(s) \propto s^d$.  To understand this dimension, consider that when the box size $s \ll R_g$, the typical potential at the edge of the box is $\sim \sqrt{s}$ (in our dimensionless units).  One can therefore imagine that the energy of donor sites within the box is chosen from a random (Gaussian) distribution with width $\sim \sqrt{s}$ and mean $\Delta/2$.  The proportion of donor sites having energy in the window $(\Delta/2, \Delta/2 + d\e)$ is therefore $\sim d\e/\sqrt{s}$.  Since there are $\sim s^3$ total sites in the box, the total number of sites within the box that are also within the energy window is $\sim s^3 d\e/\sqrt{s}$, and therefore $M(s) \sim s^{5/2} d\e$.  One can therefore conclude that sites with energy near $\pm \Delta/2$ have a fractal dimension $d = 5/2$ at length scales $s \ll R_g$.   Notice in particular that this result for $M(s)$ matches the one derived in the previous paragraph for $s \gg R_g$ when $s \sim R_g \sim \Delta^2$.

\subsubsection{Fractal dimension of sites with energy close to zero}

We now turn our attention to those sites with energy very close to zero.  In particular, let us consider the set of sites with energy $\e \in (-\e_0, \e_0)$, where $\e_0 \ll \Delta$; we refer to this set as the ``$\e_0$ set".   Constraints on the value of $\e_0$ will be discussed below. The spatial arrangement of sites within the $\e_0$ set is influenced by the structure of the random potential (including the process of nonlinear screening) and by the correlations between filled and empty sites implied by the ES criterion [Eq.\ (\ref{eq:EScrit})].  To derive the effective spatial dimension of these sites, let us first note that for the $\e_0$ set there are two relevant length scales (in addition to the lattice constant).  On the one hand, at length scales $s \gg R_g$, the random potential loses its correlations by the process of nonlinear screening, and sites within the set must be three-dimensionally distributed.  One can therefore say that at distances $s \gg R_g$ the behavior of the function $M(s)$ is given by $M(s) \sim [\int_{-\e_0}^{\e_0} g(\e) d\e] s^3 \sim \e_0 g(\e_0) s^3$.

While the length $R_g$ is associated with potential fluctuations of order $\Delta$, there is another, much shorter length scale $R_0 \sim \e_0^2$ associated with fluctuations of order $\e_0$.  That is, over scales $s \ll R_0$ the potential does not develop sufficiently to bring donor or acceptor sites from energy $\sim \e_0$ to the Fermi level.  One can therefore expect that at distances $s \ll R_0$ the function $M(s)$ behaves equivalently as for the sites that have energy far from zero, and $M(s) \sim \e_0 s^{5/2}$.

At intermediate length scales, $R_0 \ll s \ll R_g$, sites within the $\e_0$ set are arranged with some fractal dimension $d$ that is to be determined.  Let us therefore write the function $M(s)$ in piecewise form
\be 
M(s) \sim 
\begin{cases}
\e_0 s^{5/2}, & s \ll R_0 \\
A s^d, & R_0 \ll s \ll R_g  \\
\e_0 g(\e_0) s^3, & s \gg R_g ,
\end{cases}
\label{eq:Mpiecewise}
\ee
where $A$ is an unknown constant to be determined.  In order to solve for $d$, let us first note that if a $d$-dimensional system having DOS $g_d(\e)$ (with units of energy per length to the power $d$) is embedded inside a 3-dimensional volume having size $R_g$, then its 3D DOS (with units of energy per volume) is $g(\e) \sim g_d(\e)/R_g^{3-d}$.  Let us then make the crucial assumption that within the $d$-dimensional space the DOS is as large as it can be, limited only by the ES bound [Eq.\ (\ref{eq:CG})].  In other words, we set $g_d(\e) \sim |\e|^{d-1}$, and therefore 
\be 
g(\e) \sim \frac{|\e|^{d-1}}{R_g^{3-d}}.
\label{eq:gesubspace}
\ee

Given this assumption, one can derive the value of the fractal dimension $d$ (and the constant $A$) by demanding continuity of the function $M(s)$ at both $s \sim R_0$ and $s \sim R_g$.  By examining Eq.\ (\ref{eq:Mpiecewise}), and substituting $R_0 \sim \e_0^2$ and $R_g \sim \Delta^2$, we get $A \sim \e_0^2$ and $d = 2$.  So we conclude that the fractal dimension of the $\e_0$ set at length scales $R_0 \ll s \ll R_g$ is $2$.

One can now ask the question: what is the largest value of $\e_0$ for which this derivation is applicable?  In particular, consider that in a volume of size $R_g$ a two-dimensional (2D) subspace contains only $\sim R_g^2$ total sites.  Sites within the $\e_0$ set must therefore comprise only $\sim 1/R_g$ of the total.  This gives a constraint $\e_0 g(\e_0) \lesssim 1/R_g$, and using Eq.\ (\ref{eq:gesubspace}) we arrive at $\e_0 \lesssim 1$. 

To summarize, our scaling analysis leads us to the following dramatic conclusion: sites with energy $|\e| \lesssim 1$ are arranged along a two-dimensional subspace over the parametrically wide range of length scales $1 \ll s \ll R_g$.

\begin{figure}[htb]
\centering
\includegraphics[width=0.48 \textwidth]{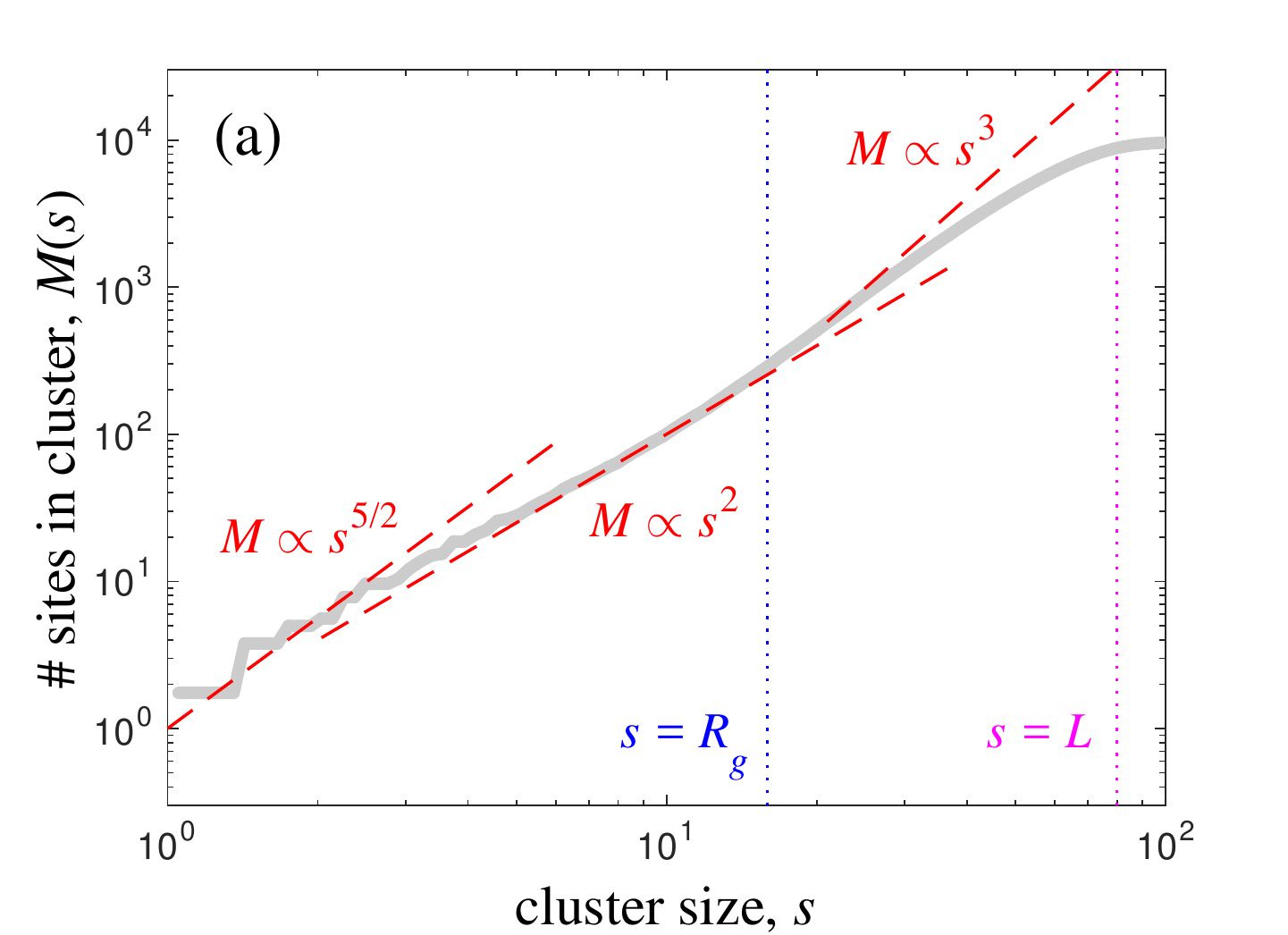}
\includegraphics[width=0.45 \textwidth]{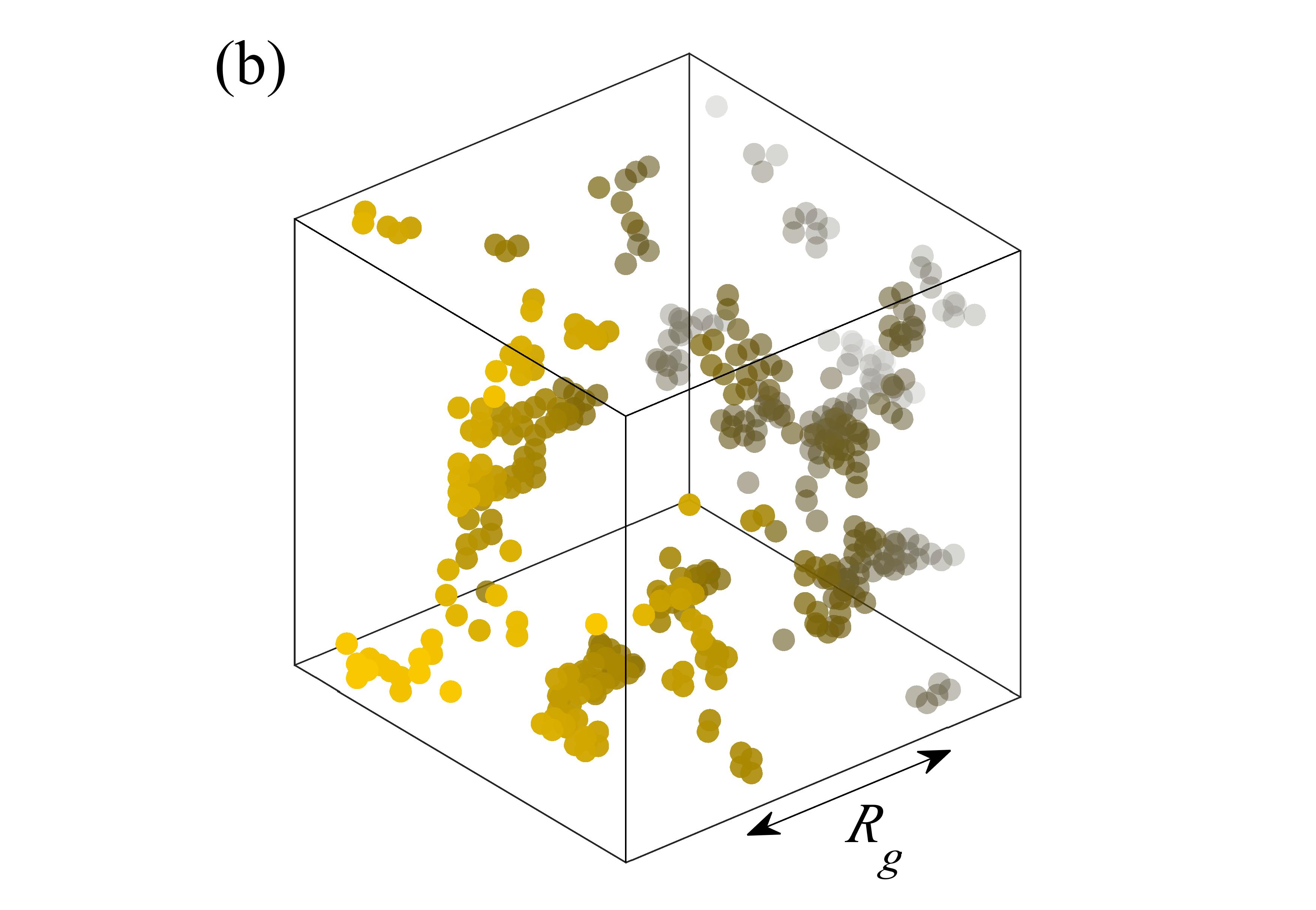}
\caption{Fractal ordering of sites with $|\e| < 1$ for a system with size $L = 80$ and gap $\Delta  = 20$.  (a) The function $M(s)$ is plotted, as measured by our numeric simulation (thick grey line) and as derived theoretically in Eq.\ (\ref{eq:Mpiecewise}) (red dashed lines).  The dashed lines use the derived values $d = 2$ and $A \sim 1$ and have no free parameters.  The vertical dotted lines indicate the length scales $R_g$ and $L$.  (b) An example of the spatial arrangement of sites with $|\e| < 1$.  Points show the locations of such sites in a typical pseudo-ground state.  (Only a subvolume of the total simulated system is shown.)  The points are colored according to their distance from the front left face of the system.}
\label{fig:fractald}
\end{figure}

This theoretical prediction can be compared directly to results from our numerical simulations, from which we can measure the function $M(s)$ for sites with energy in the band $(-1, 1)$ (see Sec.\ \ref{sec:fractalnumerics}).  The results are shown in Fig.\ \ref{fig:fractald}(a) for a system of size $L = 80$ and gap $\Delta = 20$, and they compare favorably with the derived function in Eq.\ (\ref{eq:Mpiecewise}).  A drawing of the typical spatial configuration of sites with $|\e| < 1$ is displayed in Fig.\ \ref{fig:fractald}(b).

\subsection{Low-energy DOS}

Using the results of the preceding subsection, one can understand the low-energy DOS as follows.  At vanishingly small energies, the DOS is constrained by the 3D Coulomb gap equation, $g(\e) \sim \e^2$.  Such a parabolic depletion of the DOS arises from the ES stability criterion, Eq.\ (\ref{eq:EScrit}), applied over distances much longer than the correlation length, $r \gg R_g$.  This relation therefore constrains the density of states only at energies $|\e| \sim 1/r \ll 1/R_g$.  

\begin{figure}[htb]
\centering
\includegraphics[width=0.48 \textwidth]{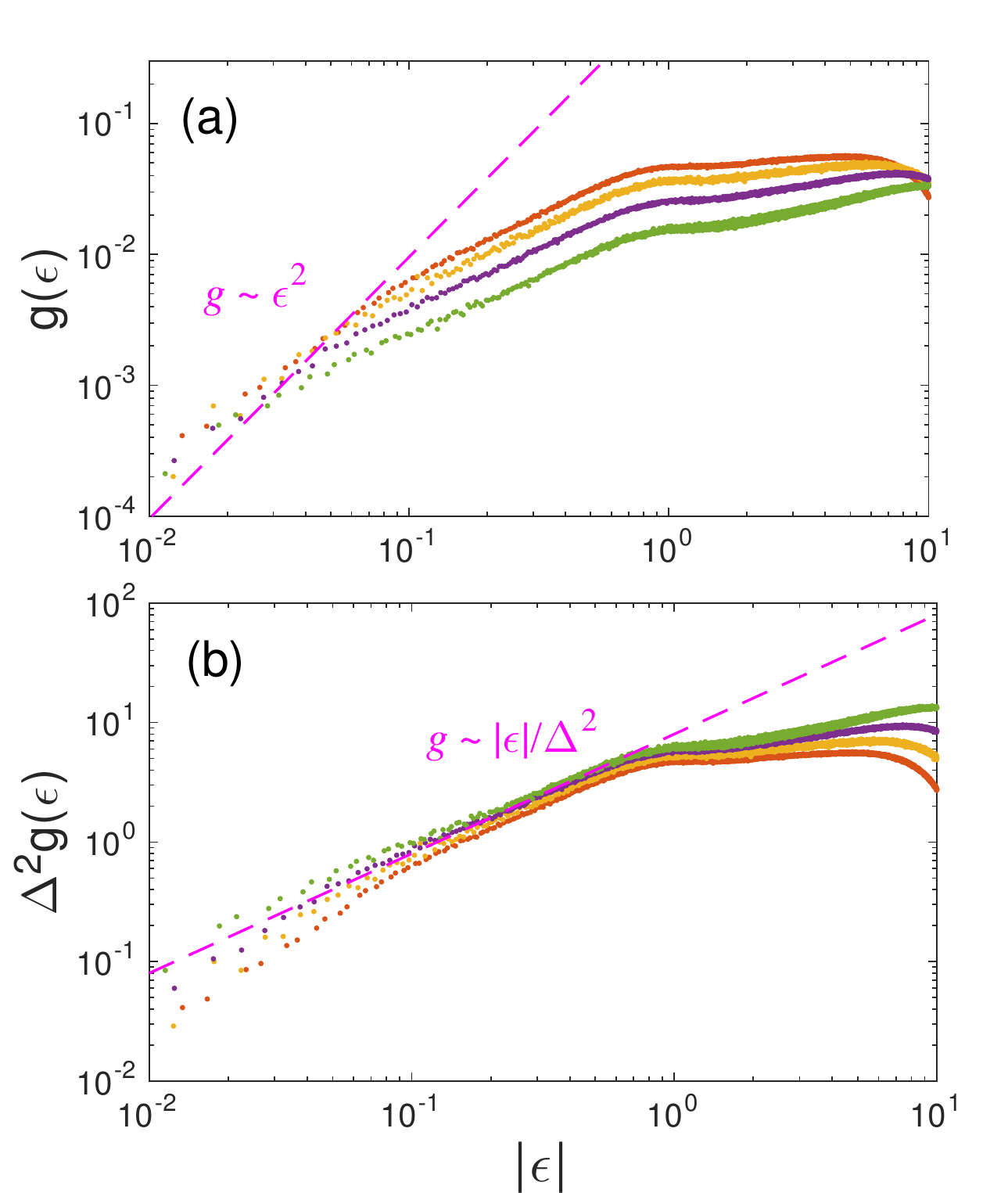}
\caption{(a) Density of states as a function of energy for different values of $\Delta$, as measured by our numeric simulations. From highest to lowest, the curves correspond to $\Delta = 10, 12, 15$, and $20$.  The dashed line shows the 3D Coulomb gap equation. (b) When the DOS is plotted as $\Delta^2 g(\e)$ versus energy, all the data collapses onto a single curve $g(\e) \sim |\e|/\Delta^2$ within the energy range $1/R_g \ll \e \ll 1$. }
\label{fig:DOSscaling}
\end{figure}

At larger energies $1/R_g \ll \e \ll 1$, the DOS is constrained by the ES criterion applied among pairs of sites with separation $r \ll R_g$; \textit{i.e.,} among sites within the same 2D cluster.  This constraint produces the DOS presented in Eq.\ (\ref{eq:gesubspace}) (with $d = 2$).  Putting these two results together gives the following result for the low-energy DOS:
\be 
g(\e) \sim 
\begin{cases}
\e^2, & \e \ll 1/R_g  \\
|\e|/R_g, & 1/R_g \ll \e \ll 1.
\end{cases}
\label{eq:gelow}
\ee
These equations coincide closely with simulation results at different values of $\Delta$, as demonstrated in Fig.\ \ref{fig:DOSscaling}.

The implications of this DOS for the conductivity are discussed in detail in the following section.  However, already from the derivation of the DOS one can understand the reason for the persistence of ES conductivity outside the regime of a parabolic DOS.  Within the range of temperatures where the activation energy $\e_a$ satisfies $1/R_g \ll \e_a \ll 1$, hopping takes place not along a set of randomly-positioned sites in three dimensions, but rather within a 2D subspace of sites having $|\e| \lesssim 1$.  The existence of this subspace greatly enhances the conductivity, and also allows the Coulomb gap to play a much more prominent role in the conductivity.

\section{Temperature-dependent conductivity}
\label{sec:conductivity}

The central result of this paper, derived in the preceding section, is that in this system the long-ranged Coulomb potential induces sites with energy $|\e| \lesssim 1$ to arrange along 2D subspaces of the sample volume.  Applying the ES stability criterion to such sites gives a DOS that varies linearly with energy.  One can therefore understand the conductivity in a relatively simple way, by adapting the Mott doctrine to account for the reduced dimensionality of the system over length scales $1 \ll r \ll R_g$.  In particular, inserting $\alpha = 1$ and $d = 2$ into Eq.\ (\ref{eq:Mottdoctrine}) gives an ES-like temperature exponent, $\ln \sigma \propto T^{-1/2}$, while a naive use of the system dimensionality $d = 3$ would give an erroneous temperature exponent $\ln \sigma \propto T^{-2/5}$. 

\begin{figure}[htb]
\centering
\includegraphics[width=0.48 \textwidth]{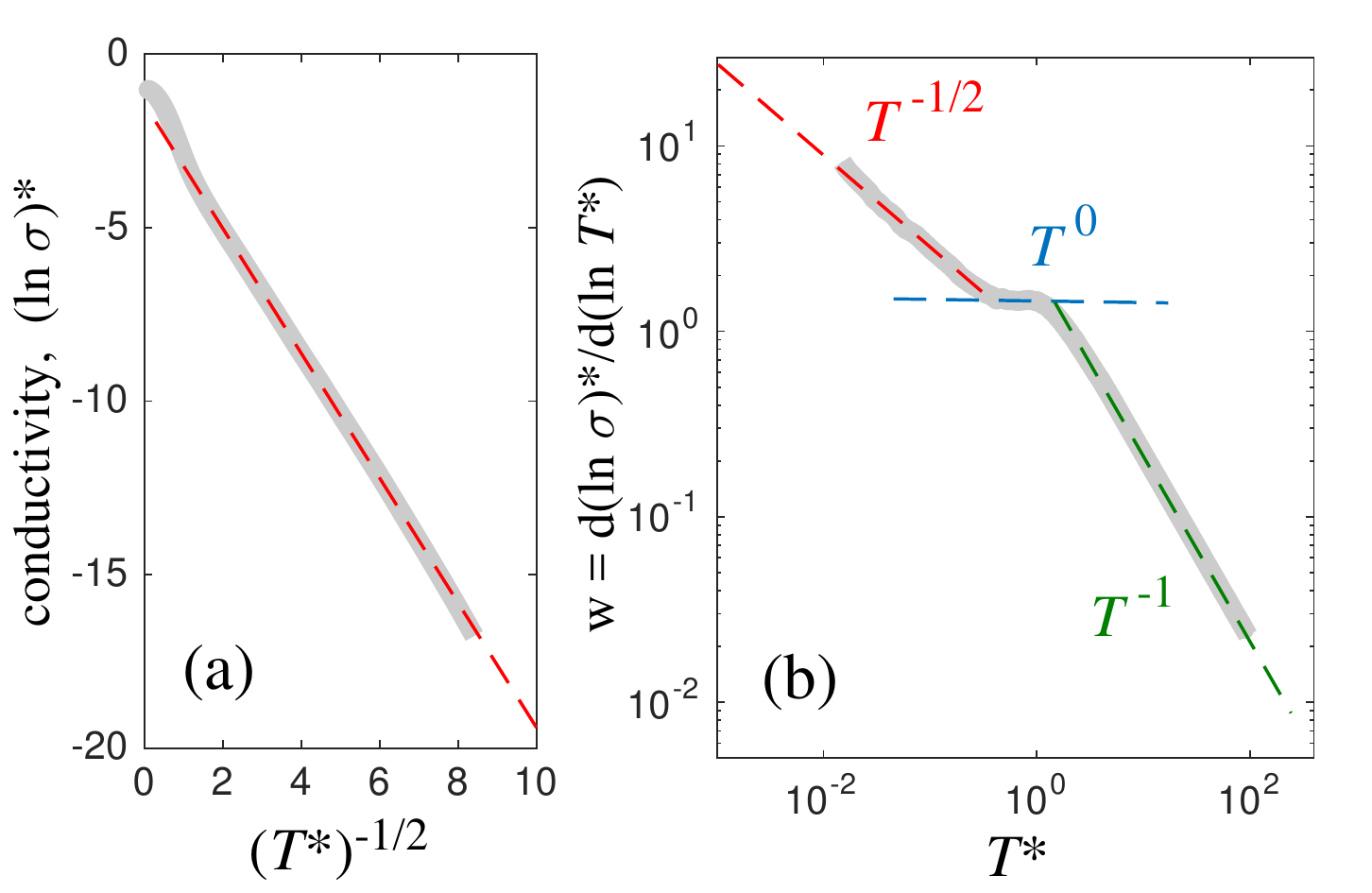}
\caption{Temperature-dependent conductivity for $\Delta = 10$, as measured by our numeric simulation. (a) The dimensionless logarithm of the conductivity, $(\ln \sigma)^*$, as a function of $(T^*)^{-1/2}$.  The thick grey curve shows our numerical result, and the dashed red line is a linear fit.  The linear dependence at $(T^*)^{-1/2} \gtrsim 1$ implies that the conductivity follows Eq.\ (\ref{eq:ES}) at all $T^* \lesssim 1$.  (b) Zabrodskii-Zinov'eva analysis of conductivity at different temperature regimes\cite{zabrodskii_influence_1975, zabrodskii_low-temperature_1984}.  The vertical axis shows the reduced activation energy $w = d(\ln \sigma)^*/d(\ln T^*)$, which has a power-law dependence on $T^*$ with an exponent that reflects the VRH mechanism. The thick grey curve is our numerical result, and dashed lines indicate the results of Eq.\ (\ref{eq:sigmapiecewise}).}
\label{fig:conductivity}
\end{figure}

Further, applying the Mott argument to this reduced-dimensional space suggests that ES conductivity persists all the way until $T^* \sim 1$.  The correctness of this conclusion is verified explicitly in Fig.\ \ref{fig:conductivity}(a), where we plot the logarithm of the conductivity against $(T^*)^{-1/2}$.  Note that the appearance of ES conductivity at relatively high temperatures $T^* \sim 1$ is in strong contrast to the result one would derive from the measured DOS using the conventional Mott doctrine.  Indeed, such a calculation would suggest that ES conductivity occurs only at extremely small temperatures, $T^* \lesssim 1/\Delta^4$, due to the tiny domain of the parabolic Coulomb gap.  

From our simulation we can also measure the numerical value of the characteristic temperature $\TES$, defined by Eq.\ (\ref{eq:ES}).  This measurement gives $\TES/(e^2/\xi) = 6.4 \pm 0.1$.  This result is consistent with the value of $\TES$ reported for 2D systems, for which numerical studies\cite{tsigankov_variable_2002} give $\TES/(e^2/\xi) \approx 6$ and a self-consistent theory\cite{levin_coulomb_1987} gives $\TES/(e^2/\xi) = 6.5$.  This agreement lends further support to our picture of hopping on fractal 2D subspaces.   In principle, at extremely small temperatures $T^* \ll 1/\Delta^4$ the typical hop length becomes much longer than $R_g$, and VRH is 3D in nature rather than 2D.  Such 3D ES conductivity is known to correspond to a different numerical value of $\TES/(e^2/\xi) \approx 2.8$.\cite{shklovskii_electronic_1984}  Unfortunately, we are unable to confirm this change in $\TES$ at low temperatures due to finite size limitations, which restrict our simulations to shorter hop lengths.

While the primary purpose of this paper is to study and explain the low temperature conductivity, it is worth commenting on the conductivity at higher temperatures as well.  In particular, one can ask about the crossover between VRH at low temperature and nearest-neighbor activated conductivity at higher temperature.  Our picture of the low-temperature conductivity is that at $T^* \ll 1$ conductivity occurs through VRH on 2D subspaces having low energy $|\e| \lesssim 1$.  As the temperature is raised to the point that $T^* \sim 1$, the typical hopping length becomes of order unity.  It is therefore natural to ask whether these 2D subspaces percolate across the system, in the nearest-neighbor sense.  If they do, then at high temperatures electrons can hop across the system using only chains of nearest-neighboring sites with $|\e| \lesssim 1$, and the corresponding activation energy will be of order unity, regardless of the value of $\Delta$.  On the other hand, if the 2D subspaces do not produce nearest-neighbor percolation across the system, then at high temperatures the current will be forced to pass through high energy sites, and the activation energy will increase with increasing $\Delta$.

\begin{figure}[htb]
\centering
\includegraphics[width=0.48 \textwidth]{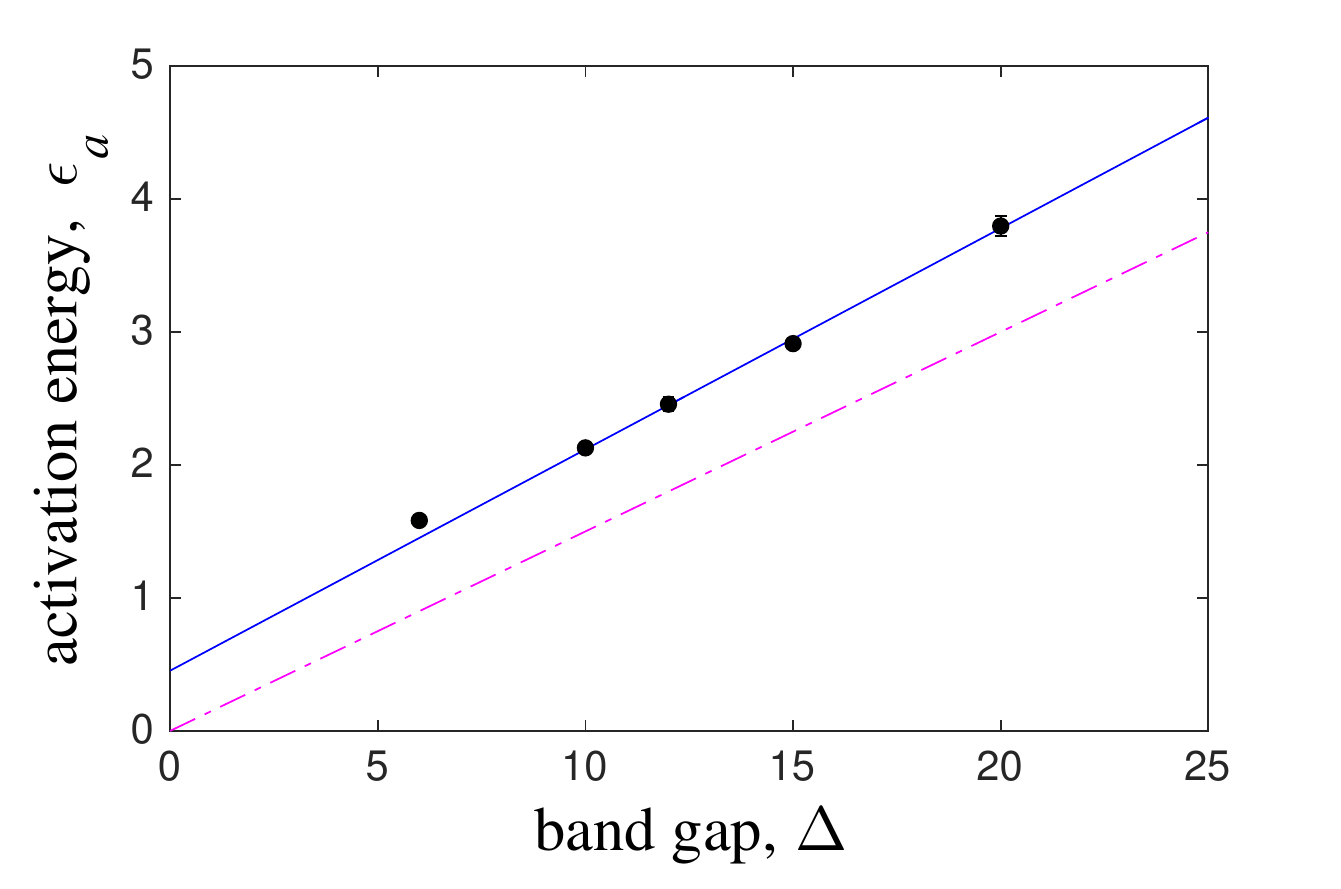}
\caption{Activation energy for nearest-neighbor hopping, $\e_a$, as a function of the band gap $\Delta$.  Black points are our numerical results, and the solid line is a linear fit.  The dash-dotted line is the result reported in Ref.\ \onlinecite{skinner_why_2012}.}
\label{fig:activation}
\end{figure}

In order to distinguish between these two possibilities, we measure the activation energy $\e_a$ for nearest-neighbor hopping as a function of $\Delta$, using the method described in Sec.\ \ref{sec:NNH}.  The result is plotted in Fig.\ \ref{fig:activation}.  Our results suggest that $\e_a$ increases linearly with $\Delta$ as $\e_a \approx 0.17 \Delta + 0.45$. This is consistent with an earlier theoretical study of the bulk transport in completely compensated topological insulators\cite{skinner_why_2012}, which studied the case $\Delta = 10$ (for a model where donors and acceptors are placed randomly, rather than on a lattice) and found that $\e_a \approx 0.15 \Delta$.

The observed increase of $\e_a$ with $\Delta$ implies that there is no nearest-neighbor percolation of 2D subspaces.  Instead, one can envision that the 3D system has many 2D subspaces embedded within it, each of size $\sim R_g$, separated by small gaps of order one lattice site [as suggested by the picture in Fig.\ \ref{fig:fractald}(b)].  When the typical hopping length is slightly larger than unity, electrons can hop across these gaps, and the activation energy is $\e_a \sim 1$.  However, if the hopping length is reduced to precisely one lattice constant, then the activation energy increases sharply to $\e_a \sim 0.17 \Delta$, which is (by our measurement) the energy required to traverse the gaps between 2D subspaces and produce nearest-neighbor percolation.  

This picture suggests that the activation energy for conductivity has an essentially discontinuous dependence on the hopping length $r$ when $r$ approaches $1$.  If one takes this picture literally, then the temperature dependence of the conductivity is given by
\be 
(\ln \sigma)^* \sim 
\begin{cases}
- (1/T^*)^{1/2}, & T^* \ll 1 \\
- 1, & 1 \ll T^* \ll 0.17 \Delta  \\
- (0.17 \Delta/T^*), & T^* \gg 0.17 \Delta .
\end{cases}
\label{eq:sigmapiecewise}
\ee
These three regimes of temperature dependence are consistent with our simulation results, as shown in Fig.\ \ref{fig:conductivity}(b).  

The intermediate temperature regime in Eq.\ (\ref{eq:sigmapiecewise}), in which $\ln \sigma$ is roughly temperature-independent, is somewhat unusual in the context of hopping transport.  (In general, our results suggest that $\sigma$ will have only a power-law dependence on temperature in the intermediate temperature range.)  It remains unclear whether such a flat temperature dependence can persist in more general models where electron sites are randomly-positioned, or whether it is an artifact of our use of a discrete lattice.  It is worth noting, however, that a similar progression of temperature exponents, including an intermediate regime where $(\ln \sigma)^* \sim 1$, has been reported experimentally\cite{wienkes_electronic_2012, wienkes_conduction_2013}.

\section{Summary and Discussion}
\label{sec:conclusion}

In this paper we have studied the problem of VRH in a simple model of a completely-compensated semiconductor, and shown that hopping conductivity is greatly enhanced at low temperatures relative to the canonical expectation using the Mott doctrine.  This enhancement arises from fractal ordering of low-energy sites, which are pushed by the long-ranged random Coulomb potential into low-dimensional subspaces having fractal dimension $d = 2$.  We have derived this fractal dimension using simple scaling arguments, which we confirm with numeric results.  The resulting picture allows us to derive theoretically the DOS and the temperature-dependent conductivity [Eqs.\ (\ref{eq:gelow}) and (\ref{eq:sigmapiecewise}), respectively].  Both results are in close agreement with numerical simulations.  Our findings represent a significant update to the theory of electronic transport in completely compensated semiconductors, which was initialized over 45 years ago\cite{shklovskii_impurity_1971, shklovskii_completely_1972}.

The most prominent experimental consequence of our results is that ES conductivity persists to a much higher temperature than the naive expectation: $T^* \sim 1$ rather than $T^* \sim 1/\Delta$.  Such enhancement has in fact been seen recently in a number of contexts.  For example, a series of works recently explored the hopping conductivity in assemblies of semiconductor nanocrystals\cite{liu_mott_2010, skinner_theory_2012, sharma_electrical_2012, chen_carrier_2014, chen_metal-insulator_2016}.  In such systems conductivity takes place through hopping of electrons between nanocrystals. Within each nanocrystal, the conduction band is split into a sequence of discrete energy shells associated with quantum confinement of electrons.  When the doping level in the assembly is such that on average an integer number of these shells is completely filled, the system resembles a completely compensated semiconductor, with the gap between quantum confinement shells playing the role of the band gap $\Delta$.  This situation was studied extensively both experimentally\cite{liu_mott_2010} and theoretically\cite{skinner_theory_2012}, and it was shown\cite{liu_mott_2010} that ES conductivity persists up to temperatures as high as $T^* \sim 2$. The DOS $g(\e)$ was also found using computer modeling\cite{skinner_theory_2012} to have a highly linear dependence on energy up to $|\e| \sim 1$, which further confirms the picture presented here.

Another prominent example of enhanced hopping conductivity has arisen in the study of 3D topological insulators (TIs).  As grown, 3D TI crystals are usually $n$-type semiconductors, and a significant experimental effort has been expended with the goal of silencing their bulk conductivity.  Such efforts usually proceed through compensation of donor electrons by acceptors, so that optimally resistive TIs are completely compensated semiconductors\cite{ren_optimizing_2011}.  However, even in the completely compensated state, the bulk conductivity is often frustratingly large experimentally, and this large conductivity has been explained in terms of the band-bending effects discussed in Sec.\ \ref{sec:DOS}.\cite{skinner_why_2012, skinner_effects_2013, borgwardt_self-organized_2016}  While much attention was paid to the unexpectedly small activation energy for conductivity at high temperatures (as examined in Fig.\ \ref{fig:activation}), these systems also display an unusually robust VRH at low temperatures.  In particular, VRH conductivity with temperature exponent close to $1/2$ has been seen to persist up to temperatures as high as $100$\,K in compensated samples of Bi$_2$Se$_3$.\cite{ren_optimizing_2011}  This is in strong contrast to the conventional expectation (as outlined in the Introduction), which instead predicts ES conductivity to end at temperatures of order $10$\,K.  This puzzling observation is explained by the theory developed here, and indeed in Bi$_2$Se$_3$ the temperature $T^* = 100$\,K corresponds to $T^* \sim 1.5$, so that this observation is completely consistent with our results.

Finally, let us remark on the generality of our results.  Most generally, one can expect our results to apply for any system that has both a finite concentration of unscreened charged impurities and a chemical potential that lies in the middle of a large energy gap.  For such systems, a large, random Coulomb potential invariably arises and produces a nontrivial spatial structure in the energy landscape (as illustrated in Fig.\ \ref{fig:bandbending}), and therefore leads to a violation of the Mott doctrine.  Of course, our model has not included the presence of ``diagonal disorder" (\emph{i.e.,} uncorrelated on-site disorder), which  generically exists with some finite amplitude $W$.  (For example, in nanocrystal arrays such disorder can arise from random variations in the nanocrystal diameter.)  However, if this diagonal disorder amplitude is small enough that $W \ll \Delta$, then its effects are overwhelmed by those of the long-ranged Coulomb potential, and our model applies.  In particular, our primary conclusion that that ES conductivity persists up to temperatures $T^* \sim 1$ remains unchanged.  On the other hand, for systems with incomplete compensation, such that either donor or acceptor type impurities predominate, the chemical potential shifts toward either the conduction or valence band edge and our theory can lose its validity.  Since the typical amplitude of the disorder potential in a system with degree of compensation $K$ is $\Gamma \sim e^2 N_i^{1/3}(1-K)^{-1/3}$,\cite{shklovskii_electronic_1984, chen_anomalously_2013} one can expect our results to apply so long as $1-K \ll 1$ and $\Gamma \gg W$.  A full analysis of the case of finite compensation and finite diagonal disorder is left for a future study.

\acknowledgments 

We are grateful to B.\ I.\ Shklovskii, G.\ Bunin, and A.\ Nahum for valuable discussions.  TC was partially supported by West Chester University Fund No. 7511001313, and BS was supported as part of the MIT Center for Excitonics, an Energy Frontier Research Center funded by the U.S. Department of Energy, Office of Science, Basic Energy Sciences under Award no. DE-SC0001088.

\bibliography{outsideCG}

\end{document}